\documentclass{article}
\usepackage{spconf,amsmath,graphicx,bm,nth,flushend}

\title{Noise robust integration for blind and non-blind reverberation time estimation}
\name{Christian Sch\"uldt$^{1,2}$ and Peter H\"andel$^1$}
\address{$ ^1$ Department of Signal Processing, ACCESS Linnaeus Centre, \\
Royal Institute of Technology KTH, SE-100 44 Stockholm, Sweden. \\
$ ^2$ Limes Audio AB, Glasbruksgatan 36, SE-116 20 Stockholm, Sweden.}

\begin{document}
\ninept
\maketitle
\begin{abstract}
The estimation of the decay rate of a signal section is an integral component of both blind and non-blind reverberation time estimation methods. Several decay rate estimators have previously been proposed, based on, e.g., linear regression and maximum-likelihood estimation. Unfortunately, most approaches are sensitive to background noise, and/or are fairly demanding in terms of computational complexity.
This paper presents a low complexity decay rate estimator, robust to stationary noise, for reverberation time estimation. Simulations using artificial signals, and experiments with speech in ventilation noise, demonstrate the performance and noise robustness of the proposed method.
\end{abstract}
\section{Introduction} \label{sec:intro}
The reverberation time of an enclosure, $T_{60}$, defined as the time required for the reverberation to decay $60$~dB, has been studied since the late \nth{19}~century. Traditionally (since the 1930s), a loudspeaker producing an interrupted noise burst is used, and the decaying signal in the enclosure is measured after the loudspeaker has gone silent, to obtain a $T_{60}$ estimate~\cite{iso3382}. Multiple measurements are typically required, due to noise fluctuations. In 1965, Schr\"oder showed that the ensemble average of decaying squared white noise is identical to a certain integral over the squared room impulse response (RIR), implying that the reverberation time could be obtained directly from the RIR, rather than from measuring the decay of multiple noise bursts~\cite{schroeder65}. The integration scheme in~\cite{schroeder65} is commonly denoted Schr\"oder backward integration, or simply backward integration.

More recently, the concept of \emph{blind} $T_{60}$ estimation has been explored. The word blind in this context means that only a reverberant recording, typically containing speech, is used for $T_{60}$ estimation, and no information about the excitation signal is available. Some methods based on machine learning have been proposed~\cite{hirsch08,xiong13}, although they tend to be computationally demanding and, obviously, require prior training. A more common approach is to directly estimate the decay rate in sections of the signal where the reverberation is dominant. Several such methods, based on direct linear regression (LR)~\cite{dumortier14,wen08,talmon13,naylor13,lebart01}, and maximum-likelihood estimation (ML)~\cite{ratnam03,lollmann08,lollmann10} have been presented, as well as methods extending the backward integration approach to the blind estimation case~\cite{prego12,vesa05}. Blind reverberation time estimation algorithms typically require, in addition to straight-forward decay rate estimation, other processing steps such as, e.g., segmentation (finding signal sections where the reverberation is dominant)~\cite{lebart01,lollmann10,prego12}, pre-whitening~\cite{keshavarz12,lopez12}, and various post-filtering methods~\cite{dumortier14,wen08,naylor13,lollmann10}. The focus of this paper is, however, solely on the decay rate estimation step, as this is typically the core of the $T_{60}$ estimation algorithm. Nevertheless, it should be emphasized that for a complete blind $T_{60}$ estimation solution, pre- and post processing methods (such as mentioned above) should naturally be used as well, and that the decay rate estimator proposed in this paper may be used in conjunction with essentially any such method(s).

Motivated by a recent comparison of blind $T_{60}$ estimation approaches~\cite{gaubitch12}, which noted that the performance of all compared estimators deteriorates in the presence of noise, this paper proposes a noise robust decay rate estimator. The proposed method is based on backward integration, requires low computational complexity, and could be used for both blind- and non-blind $T_{60}$ estimation. Monte-Carlo simulations with artificial signals, as well as experiments with speech signals corrupted by ventilation noise, are used to verify the performance.

\section{Decay rate estimation}
A common model to describe the RIR, assuming a diffusive sound field and a source-microphone distance greater than the critical distance~\cite{kuttruff00}, was introduced by Polack~\cite{polack88} as
\begin{equation} \label{eqn:rir}
	f(t) = v(t) e^{-t \rho} + d(t) \qquad t \ge 0,
\end{equation}
where $t$ is the (continuous) time, $\rho > 0$ is the decay rate, and $v(t)$ and $d(t)$ are two wide-sense stationary uncorrelated random processes, representing the (decaying) reverberation and the background noise, respectively. The decay rate $\rho$ is related to the reverberation time as
\begin{equation} \label{eqn:RT60}
	T_{60} = \frac{3}{\rho \log_{10}(e)} \approx \frac{6.91}{\rho}.
\end{equation}
In non-blind $T_{60}$ estimation, the decay rate can be estimated from the RIR directly (using any type of decay rate estimator), to obtain the reverberation time. In the blind case, it can be assumed that the reverberated signal (either in full-band or in frequency sub-bands) locally adheres to a similar exponential model~\cite{wen08,ratnam03}. The same type of decay rate estimation can, thus, be used for both blind and non-blind $T_{60}$ estimation.

In the following, techniques for decay rate estimation, commonly used in the context of reverberation time estimation, are briefly described.

\subsection{Linear regression} \label{sec:lr}
Perhaps the most basic decay rate estimator is based on simply fitting a straight line to $\ln f^2(t)$, assuming that the influence of the background noise $d(t)$ is negligible. The slope of the line then corresponds to the decay rate. This linear regression approach has been used for decay rate estimation in many blind reverberation time estimators; see, e.g.,~\cite{dumortier14,wen08,talmon13,naylor13,lebart01}. It has been shown that in a noiseless scenario, this estimator is unbiased and achieves an estimation variance approximately $4$~dB above the Cram\'{e}r-Rao lower bound (CRB)~\cite{schuldt14}. In the case of noise, however, the estimator suffers from significant bias~\cite{schuldt14}.

\subsection{Non-linear regression} \label{sec:nlr}
Unfortunately, in many cases the background noise is not negligible, meaning that the ensemble mean of $\ln f^2(t)$ will not adhere to straight line. Approaches for non-linear regression, where the influence of the noise is taken into account, has been proposed for non-blind reverberation time estimation~\cite{xiang95, karjalainen02}. However, non-linear regression requires numerical optimization and is thus demanding, especially compared to simple linear regression, in terms of computational complexity.

\subsection{Backward integration}
Backward integration has been used extensively for non-blind reverberation time estimation, and is also part of the ISO~3382 standard~\cite{iso3382}. The concept, as introduced by Schr\"oder~\cite{schroeder65}, does not rely on the Polack model assumption, although, under the assumption of the Polack model~(\ref{eqn:rir}), backward integration can also be used for blind reverberation time estimation (see, e.g.,~\cite{prego12,vesa05}), owing to the fact that the integral of an exponentially decaying waveform also is exponentially decaying. 
%
Integrating~(\ref{eqn:rir}) significantly reduces the variance, and the reverberation time can then be estimated, as in Section~\ref{sec:lr}, by fitting a line to the logarithm of the integrated curve. However, practical problems exist due to background noise and the use of a finite upper integration limit. This causes a bending of the tail of the integrated curve~\cite{schuldt14,schuldt15,morgan97}, resulting in a decay rate estimation bias. Methods for handling the background noise problem, through, e.g., subtraction of a background noise estimate~\cite{chu78}, or by careful tuning of the line fitting and integration limits~\cite{morgan97}, have been proposed. However, it has recently been shown that tuning of the backward integration parameters is non-trivial, as the optimal parameter selection depends (among other things) on the actual decay rate~\cite{schuldt15}, and that this is especially critical when the background noise is strong, which typically can be the case for blind reverberation time estimation~\cite{schuldt15}.

\subsection{Maximum-likelihood estimation} \label{sec:ml}
In~\cite{lollmann08}, a noise robust ML decay rate estimator was presented, extending the previous work in~\cite{ratnam03}, which did not take background noise into account. Assuming a time discrete version of~(\ref{eqn:rir}),
\begin{equation}
	f_d(n) = v_d(n) e^{-n \rho_d} + d_d(n),
\end{equation}
where $n \ge 0$ is the sample index, and assuming that $v_d(n)$ and $d_d(n)$ are two uncorrelated i.i.d. Gaussian signals with zero mean and with variances $\sigma^2_{v,d}$ and $\sigma^2_{d,d}$, respectively, gives the log-likelihood function for $N$ observations ($n=\{0,\cdots,N-1\})$ of $f_d(n)$ as~\cite{lollmann08}
\begin{align} \label{eqn:loglikelihood}
	& -\frac{1}{2} \Bigg( \ln(2 \pi) + \sum_{n=0}^{N-1} \ln(\sigma_{v,d}^2 e^{-2 n \rho_d} + \sigma_{d,d}^2 ) \nonumber \\
	& + \sum_{n=0}^{N-1} \frac{f_d^2(n)}{\sigma_{v,d}^2 e^{-2 n \rho_d} + \sigma_{d,d}^2} \Bigg).
\end{align}
Finding the maximum of~(\ref{eqn:loglikelihood}), with respect to $\sigma_{v,d}$, $\sigma_{d,d}$ and $\rho_d$ allows estimation of the decay rate, and thus the reverberation time. The maximum of~(\ref{eqn:loglikelihood}) can be found directly through, e.g., expectation-maximization~\cite{lollmann08}, although this is a fairly demanding procedure in terms of computational complexity. Another approach, denoted Hybrid MLN~\cite{schuldt14}, initially estimates $\sigma_{v,d}$ from the first $N_L$ of the $N$ observations, where it is assumed that the noise can be negligible, and then uses this information, together with a separately obtained estimate of $\sigma_{d,d}$, to find the decay rate $\rho$. (For a more comprehensive description of Hybrid MLN, the reader is referred to~\cite{schuldt14}.)

Nevertheless, the disadvantage of these ML approaches is that they all lead to transcendental equation that have to be solved numerically. A low complexity approach has been presented in~\cite{schuldt13}, using the assumption of $v_d(n)$ having Laplacian distribution, and using polynomial approximation for root finding. However, this method does not consider any noise.

\section{Proposed decay rate estimator}
The objective of the proposed decay rate estimator is to avoid the well-know backward integration tail problem~\cite{schuldt15,morgan97}, while incorporating noise robustness. This is achieved by intrinsically modeling the bend of the tail, caused by both the finite upper integration as well as the noise, similar to, e.g.,~\cite{xiang95}. However, the proposed method is based on \emph{successive integration}~\cite{halmer04},
yielding a low complexity estimator, in contrast to~\cite{xiang95}, where a computationally demanding iterative approach is used. For the sake of simplicity, the backward integration is altered so that the integration range is between $0$ and $t$, instead of between $t$ and $\infty$. The principal idea is, however, the same. Practical issues regarding, e.g., the computation of the integrals and the computational complexity, are discussed in Section~\ref{sec:practical}.

By assuming the Polack model in~(\ref{eqn:rir}), and that the random processes $v(t)$ and $d(t)$ are zero mean, wide-sense stationary and uncorrelated, the (ensemble) mean of the integrated curve can be written as
\begin{equation} \label{eqn:bi2}
	\int_0^t E\{ f^2(\tau) \} \, \mathrm{d}\tau = t \sigma^2_d + \frac{\sigma^2_v}{2 \rho} (1 - e^{-2 t \rho}),
\end{equation}
where $\sigma^2_d = E\{ d^2(t) \}$ and $\sigma^2_v = E\{ v^2(t) \}$, and $E\{\cdot\}$ denotes expected value. In many cases, $\sigma^2_d$ can be estimated separately; in the blind estimation case, one could estimate $\sigma^2_d$ in periods of silence~\cite{lollmann08}, and in the non-blind estimation case, one could simply increase the number of estimated RIR coefficients sufficiently so that the last samples are guaranteed to contain only $d(n)$. If $\sigma^2_d$ is estimated separately, the estimate can simply be subtracted from~(\ref{eqn:bi2}), i.e.,
\begin{align} \label{eqn:bin}
	g(t) &= \int_0^t E\{ f^2(\tau) \} \, \mathrm{d}\tau - t \hat{\sigma}^2_d \nonumber \\
	& = t ( \sigma^2_d - \hat{\sigma}^2_d ) + \frac{\sigma^2_v}{2 \rho} (1 - e^{-2 t \rho}),
\end{align}
and the noise influence can thus be neglected if $\hat{\sigma}^2_d \approx \sigma^2_d$, yielding a noise compensated signal $g(t)$. (Note the similarity to the noise power subtraction in~\cite{chu78}, although in that case linear fitting to the backward integrated curve formed the basis of the decay rate estimator.) The method of successive integration~\cite{halmer04} is also based on the fact that the integral of an exponentially decaying waveform is exponentially decaying as well. Hence, integrating~(\ref{eqn:bin}) again according to
\begin{align}
	\int_0^t g(\tau) \, \mathrm{d}\tau \nonumber & = \frac{\sigma^2_v}{2 \rho} \Big(t - \frac{1}{2 \rho} \big( 1 - e^{-2 t \rho}\big) \Big) \nonumber \\
	& = \frac{1}{2 \rho} \left( \sigma^2_v t + g(t) \right),
\end{align}
gives that $g(t)$ can be written as a function of its own integral, i.e.,
\begin{equation} \label{eqn:si}
	g(t) = \left( \sigma^2_v t - 2 \rho \int_0^t g(\tau) \, \mathrm{d}\tau \right).
\end{equation}
It can then be seen that a simple two dimensional function
\begin{equation} \label{eqn:si_hat}
	\hat{g}(t,s) = \alpha_0 t + \alpha_1 s + \alpha_2,
\end{equation}
can be fitted to~(\ref{eqn:si}), where the parameter $\alpha_2$ is added to somewhat lessen the influence of modeling errors (i.e., if the model in~(\ref{eqn:rir}) is not entirely correct). Given a total of $N \ge 3$ measurements of $g(t)$ at times $\{t_0, t_1, \cdots, t_{N-1}\}$, the unknown parameters $\alpha_0$, $\alpha_1$ and $\alpha_2$ can then be found through a standard least-squares approach, i.e.,
\begin{equation} \label{eqn:ls}
	\min \sum_{n=0}^{N-1} \Big( g(t_n) - \hat{g}\big(t_n, g_I(t_n) \big) \Big)^2,
\end{equation}
where $g_I(t_n) = \int_0^{t_n} g(\tau) \, \mathrm{d}\tau$. The minimum squared error is obtained by setting the partial derivatives of the expression in~(\ref{eqn:ls}) with respect to $\alpha_0$, $\alpha_1$ and $\alpha_2$ to zero, and solving for the unknown parameters. This gives the linear system
\begin{equation} \label{eqn:le}
	\begin{pmatrix}
		\displaystyle \langle \bm{t}, \bm{t} \rangle & \displaystyle \langle \bm{t}, \bm{g}_I \rangle & \displaystyle \langle \bm{t}, \bm{1} \rangle \\
		\displaystyle \langle \bm{g}_I, \bm{t} \rangle & \displaystyle \langle \bm{g}_I, \bm{g}_I \rangle & \displaystyle \langle \bm{g}_I, \bm{1} \rangle \\
		\displaystyle \langle \bm{1}, \bm{t} \rangle & \displaystyle \langle \bm{1}, \bm{g}_I \rangle & \displaystyle \langle \bm{1}, \bm{1} \rangle
	\end{pmatrix}
	\begin{pmatrix}
		\displaystyle \alpha_0 \\
		\displaystyle \alpha_1 \\
		\displaystyle \alpha_2
	\end{pmatrix}
	=
	\begin{pmatrix}
		\displaystyle \langle \bm{t}, \bm{g} \rangle \\
		\displaystyle \langle \bm{g}_I, \bm{g} \rangle \\
		\displaystyle \langle \bm{1}, \bm{g} \rangle
	\end{pmatrix},
\end{equation}
where $\langle \cdot, \cdot \rangle$ denotes inner product, $\bm{t} = [ t_0, t_1, \cdots, t_{N-1} ]^T$, $\bm{g}_I = [ g_I(t_0), g_I(t_1), \cdots, g_I(t_{N-1}) ]^T$, $\bm{g} = [ g(t_0), g(t_1), \cdots, \\ g(t_{N-1}) ]^T$ and $\bm{1} = [ 1, 1, \cdots, 1 ]^T$. Solving~(\ref{eqn:le}) for $\alpha_0$, $\alpha_1$ and $\alpha_2$ then gives the estimated decay rate as $\hat{\rho} = - \alpha_1 / 2$ (compare~(\ref{eqn:si}) with~(\ref{eqn:si_hat})). Hence, an estimate of $\rho$ is obtained directly, and the bending of the integrated curve (see, e.g.,~\cite{schuldt14,morgan97}) requires no special attention.

\subsection{Practical considerations} \label{sec:practical}
It should be noted that one typically has constant time interval between the measurements of $g(t)$, meaning that the time instances effectively can be written as $\bm{t} = [0, 1, 2, \cdots, N-1]^T$. This, in turn, means that some of the inner products can be calculated directly, e.g., $\langle \bm{t}, \bm{t} \rangle = N (N-1) (2 N - 1) / 6$. The other inner products in~(\ref{eqn:le}) can be obtained using a total of $3 N$ multiplications (note that, e.g., $\langle \bm{g}_I, \bm{1} \rangle$ can be calculated through a cumulative sum). When calculating $g(t)$ in~(\ref{eqn:bin}), the measured values of $f^2(t)$ are used, instead of the ensemble average. This can be done owing to the variance reducing effect of the integration. If integral trapezoid approximation is used (see~\cite{halmer04} for details on how this affects the estimation performance), $N$ multiplications (the squaring of $f(t)$) are required to calculate the vectors $\bm{g}$ and $\bm{g}_I$. This gives a total of $4 N$ multiplications required for the linear equation system in~(\ref{eqn:le}). The multiplications required to solve the system can be neglected since $N$ is typically large ($>100$). In comparison, the method in~\cite{schuldt14}, denoted hybrid MLN, requires $R ( 3 (N + 1) + N_L + 1)$ multiplications, where $R$ is the number of recursions for numerical root finding and $N_L \le N$ is the length of the window used for estimating $\sigma^2_v$. Hence, hybrid MLN requires approximately $R$ times more multiplications ($R = 5$ was used in~\cite{schuldt14}) in total, compared to the method proposed here.


\begin{figure}[t]
	\centering
	\includegraphics[width=8.6cm]{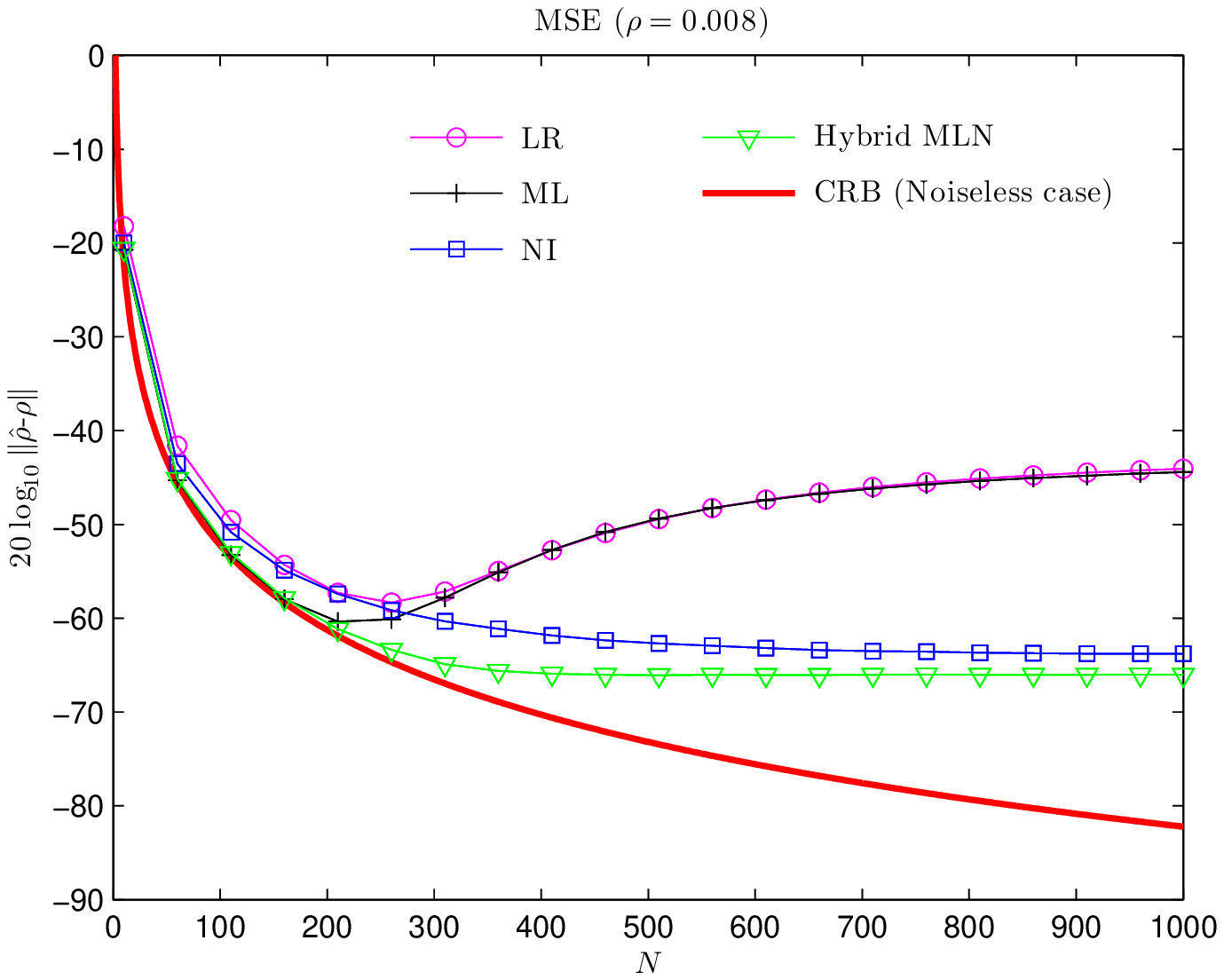} \vspace{-3mm}
	\caption{MSE of the estimators for $\rho = 0.008$ and varying $N$, in the case of i.i.d.\ Gaussian signals.}
	\label{fig:fig1}
	\vspace{5mm}
	\centering
	\includegraphics[width=8.6cm]{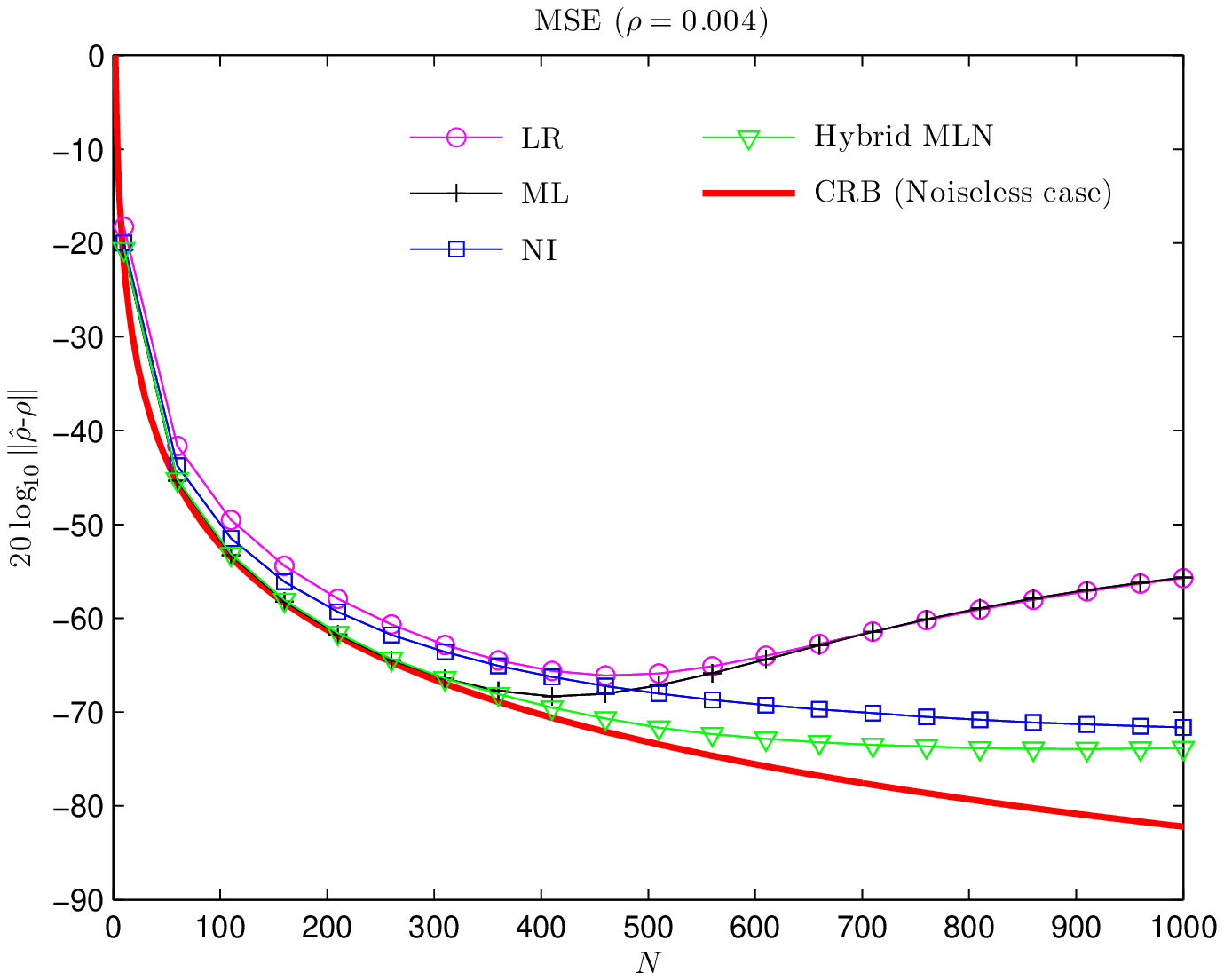} \vspace{-3mm}
	\caption{MSE of the estimators for $\rho = 0.004$ and varying $N$, in the case of i.i.d.\ Gaussian signals.}
	\label{fig:fig2}
\end{figure}

\section{Simulations} \label{sec:simulations}
\begin{figure}[t]
	\centering
	\includegraphics[width=8.6cm]{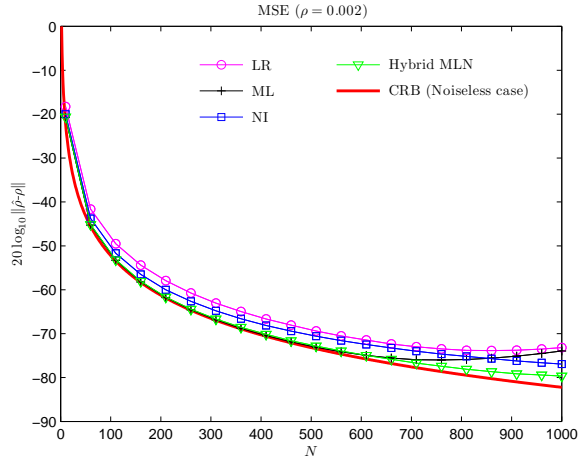} \vspace{-3mm}
	\caption{MSE of the estimators for $\rho = 0.002$ and varying $N$, in the case of i.i.d.\ Gaussian signals.}
	\label{fig:fig3}
\end{figure}

Simulations were conducted to evaluate the proposed algorithm, here denoted NI (noise robust integration). First, a discrete version of the Polack model in~(\ref{eqn:rir}) was considered, where both random processes were zero-mean i.i.d.\ Gaussian with variances $\sigma^2_v = 1$ and $\sigma^2_b = 0.01$, respectively. Three different decay rates were used, $\rho = \{0.008, 0.004, 0.002\}$ (corresponding to $T_{60}$ reverberation times of approximately $0.1$, $0.2$ and $0.4$~seconds for a sample rate $F_s$ of $8$~kHz). The decay rate was estimated with each considered method: LR (linear regression, see Section~\ref{sec:lr}), ML that does not take the noise into account~\cite{ratnam03}, hybrid MLN~\cite{schuldt14} and the proposed NI, for a observation window lengths of up to $N = 1000$ samples ($125$~ms for $F_s = 8$~kHz). It was assumed that the noise level~$\sigma^2_b$ was known for both noise robust methods NI and hybrid MLN. A total of $100 000$ Monte-Carlo simulations were performed for each of the three decay rates, and the results are shown in~Fig.~\ref{fig:fig1},~Fig.~\ref{fig:fig2} and~Fig.~\ref{fig:fig3}, respectively. The CRB (see~\cite{schuldt14}) is also shown for reference.

From the figures, it can be seen that the proposed NI exhibits lower estimation mean square error (MSE) than the LR approach for all $N$, and that it indeed is robust to the noise. Moreover, the MSE of NI is only slightly above that of that of the hybrid MLN, while having significantly lower computational complexity.

\begin{figure}[t]
	\centering \vspace{-2.4mm}
	\includegraphics[width=7.5cm,height=11.1cm]{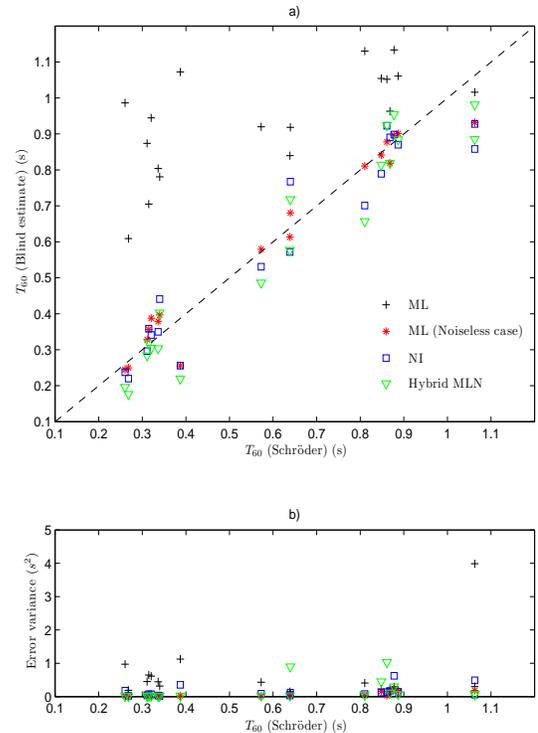} \vspace{-8mm}
	\caption{a) Median of estimated $T_{60}$ for the estimators, in the case of speech and ventilation noise. Schr\"oder's (non-blind) method is used as ground-truth. The dashed diagonal line is shown for reference. b) Error variance.}
	\label{fig:real}
\end{figure}

\subsection{Experiments with recorded RIRs and speech signals}
To test the performance in a more realistic scenario, the proposed NI method was also evaluated using a speech file, with sampling frequency $8$~kHz, containing utterances from different speakers (both male and female) recorded in an anechoic room. The speech was convoluted with $20$ different RIRs, measured in rooms with different reverberation time. The RIRs were taken from the AIR database~\cite{jeub09} and downsampled to $8$~kHz. Recorded ventilation noise was added to the reverberant speech (except for the benchmark method denoted ``ML (Noiseless case)'', in which case no noise was added), resulting in speech-to-noise ratios ranging between $9$~dB and $17$~dB, depending on the RIR. A total of $20$ different segments of varying length, containing dominating reverberant decaying power, were manually selected from the speech file, and the reverberation time was estimated from each segment. Fig.~\ref{fig:real}~a) shows the median estimated reverberation time over the $20$ speech segments for the compared methods, plotted against the reverberation time obtained by Schr\"oder's method~\cite{schroeder65} for the $20$ different RIRs, here used as ground-truth. Fig.~\ref{fig:real}~b) shows the respective error variances. It can clearly be seen that the ML approach performs well in absence of noise, but lacks noise robustness, especially when $T_{60}$ is low. (The performance of LR is very similar, although not shown in order to improve the readability of the figure.) On the other hand, NI and hybrid MLN manage to fairly accurately estimate the reverberation time, despite the noise. The performance of both these methods appear fairly equal (in accordance to what is illustrated in~Fig.~\ref{fig:fig1},~Fig.~\ref{fig:fig2} and~Fig.~\ref{fig:fig3}) also in this case, and the significantly lower computational complexity of NI (see section~\ref{sec:practical}) should be kept in mind.

It should, however, be emphasized that how the segmentation is done affects the performance. For example, the ML approach would benefit from having the segmentation step be very careful about including noise. The noise robust methods NI and hybrid MLN, on the other hand, do not suffer from degraded performance by including noise in the segments, hence relaxing the segmentation requirements. It should also be noted that the proposed method would probably facilitate blind $T_{60}$ approaches that do not rely on explicit segmentation, such as, e.g.,~\cite{dumortier14,wen08,keshavarz12}.



\section{Conclusion}
A noise robust decay rate estimator requiring low computational complexity was proposed for blind and non-blind $T_{60}$ estimation. Simulations using artificial signals, as well as experiments with speech signals convoluted with RIRs measured in rooms with different reverberation time and added recorded ventilation noise, demonstrate that the method indeed is robust to the noise, while having performance similar to the non-robust decay rate estimator in the absence of noise.

\bibliographystyle{IEEEbib}
\bibliography{refs}

\begin{thebibliography}{10}

\bibitem{iso3382}
ISO 3382-2,
\newblock ``Acoustics - {M}easurement of room acoustic parameters - {P}art 2:
  Reverberation time in ordinary rooms,'' International Standards Organization,
  Geneva, Switzerland, 2008.

\bibitem{schroeder65}
M.~R. Schr{\"o}der,
\newblock ``New method of measuring reverberation time,''
\newblock {\em The Journal of the Acoustical Society of America}, vol. 37, no.
  3, pp. 409--412, 1965.

\bibitem{hirsch08}
H-G. Hirsch and H.~Finster,
\newblock ``A new approach for the adaptation of {HMM}s to reverberation and
  background noise,''
\newblock {\em Speech Communication}, vol. 50, no. 3, pp. 244--263, 2008.

\bibitem{xiong13}
F.~Xiong, S.~Goetze, and B.T. Meyer,
\newblock ``Blind estimation of reverberation time based on spectro-temporal
  modulation filtering,''
\newblock in {\em Proceedings of the IEEE International Conference on
  Acoustics, Speech and Signal Processing (ICASSP)}, May 2013, pp. 443--447.

\bibitem{dumortier14}
B.~Dumortier and E.~Vincent,
\newblock ``Blind {RT60} estimation robust across room sizes and source
  distances,''
\newblock in {\em Proceedings of the IEEE International Conference on
  Acoustics, Speech and Signal Processing (ICASSP)}, May 2014.

\bibitem{wen08}
J.~Y. Wen, E.~A. Habets, and P.~A. Naylor,
\newblock ``Blind estimation of reverberation time based on the distribution of
  signal decay rates,''
\newblock in {\em Proceedings of the IEEE International Conference on
  Acoustics, Speech and Signal Processing (ICASSP)}, March 2008, pp. 329--332.

\bibitem{talmon13}
R.~Talmon and E.~A.~P. Habets,
\newblock ``Blind reverberation time estimation by intrinsic modeling of
  reverberant speech,''
\newblock in {\em Proceedings of the IEEE International Conference on
  Acoustics, Speech and Signal Processing (ICASSP)}, May 2013, pp. 156--160.

\bibitem{naylor13}
J.~Eaton, N.~D. Gaubitch, and P.~A. Naylor,
\newblock ``Noise-robust reverberation time estimation using spectral decay
  distributions with reduced computational cost,''
\newblock in {\em Proceedings of the IEEE International Conference on
  Acoustics, Speech and Signal Processing (ICASSP)}, May 2013, pp. 161--165.

\bibitem{lebart01}
K.~Lebart, J.~M. Boucher, and P.~N. Denbigh,
\newblock ``A new method based on spectral subtraction for speech
  dereverberation,''
\newblock {\em Acta Acustica united with Acustica}, vol. 87, no. 3, pp.
  359--366, May/June 2001.

\bibitem{ratnam03}
R.~Ratnam, D.~L. Jones, B.~C. Wheeler, W.~D. O'Brien, C.~R. Lansing, and A.~S.
  Feng,
\newblock ``Blind estimation of reverberation time,''
\newblock {\em The Journal of the Acoustical Society of America}, vol. 114, no.
  5, pp. 2877--2892, Nov. 2003.

\bibitem{lollmann08}
H.~W. L{\"o}llmann and P.~Vary,
\newblock ``Estimation of the reverberation time in noisy environments,''
\newblock in {\em Proceedings of the International Workshop for Acoustic Echo
  Cancellation and Noise Control (IWAENC)}, Sept. 2008, vol.~4.

\bibitem{lollmann10}
H.~W. L{\"o}llmann, E.~Yilmaz, M.~Jeub, and P.~Vary,
\newblock ``An improved algorithm for blind reverberation time estimation,''
\newblock in {\em Proceedings of the International Workshop for Acoustic Echo
  Cancellation and Noise Control (IWAENC)}, Aug. 2010.

\bibitem{prego12}
T.~d.~M.~Prego, A.~A. de~Lima, S.~L. Netto, B.~Lee, A.~Said, R.~W. Schafer, and
  T.~Kalker,
\newblock ``A blind algorithm for reverberation-time estimation using subband
  decomposition of speech signals,''
\newblock {\em The Journal of the Acoustical Society of America}, vol. 131, pp.
  2811--2816, 2012.

\bibitem{vesa05}
S.~Vesa and A.~Harma,
\newblock ``Automatic estimation of reverberation time from binaural signals,''
\newblock in {\em Proceedings of the IEEE International Conference on
  Acoustics, Speech and Signal Processing (ICASSP)}, March 2005, vol.~3, pp.
  281--284.

\bibitem{keshavarz12}
A.~Keshavarz, S.~Mosayyebpour, M.~Biguesh, T.A. Gulliver, and M.~Esmaeili,
\newblock ``Speech-model based accurate blind reverberation time estimation
  using an {LPC} filter,''
\newblock {\em IEEE Transactions on Audio, Speech, and Language Processing},
  vol. 20, no. 6, pp. 1884--1893, 2012.

\bibitem{lopez12}
N.~L{\'o}pez, Y.~Grenier, and I.~Bourmeyster,
\newblock ``Low variance blind estimation of the reverberation time,''
\newblock in {\em Proceedings of the International Workshop for Acoustic Echo
  Cancellation and Noise Control (IWAENC)}, Sept. 2012.

\bibitem{gaubitch12}
N.~D. Gaubitch, H.~W. L{\"o}llmann, M.~Jeub, T.~H. Falk, P.~A. Naylor, P.~Vary,
  and M.~Brookes,
\newblock ``Performance comparison of algorithms for blind reverberation time
  estimation from speech,''
\newblock in {\em Proceedings of the International Workshop for Acoustic Echo
  Cancellation and Noise Control (IWAENC)}, Sept. 2012.

\bibitem{kuttruff00}
H.~Kuttruff,
\newblock {\em Room Acoustics},
\newblock Taylor {\&} Francis, 4th ed. edition, Oct. 2000.

\bibitem{polack88}
J-D. Polack,
\newblock {\em La transmission de l'energie sonore dans les salles},
\newblock Ph.D. thesis, Le Mans, 1988.

\bibitem{schuldt14}
C.~Sch{\"u}ldt and P.~H{\"a}ndel,
\newblock ``Decay rate estimators and their performance for blind reverberation
  time estimation,''
\newblock {\em IEEE Transactions on Audio, Speech, and Language Processing},
  vol. 22, no. 8, pp. 1274--1284, Aug. 2014.

\bibitem{xiang95}
N.~Xiang,
\newblock ``Evaluation of reverberation times using a nonlinear regression
  approach,''
\newblock {\em The Journal of the Acoustical Society of America}, vol. 98, pp.
  2112--2121, 1995.

\bibitem{karjalainen02}
M.~Karjalainen, P.~Ansalo, A.~M{\"a}kivirta, and T.~Peltonen,
\newblock ``Estimation of modal decay parameters from noisy response
  measurements,''
\newblock {\em Journal of the Audio Engineering Society}, vol. 50, no. 11, pp.
  867--878, 2002.

\bibitem{schuldt15}
C.~Sch{\"u}ldt and P.~H{\"a}ndel,
\newblock ``On implications of the {ISO 3382} backward integration method for
  automated decay rate estimation,''
\newblock {\em Journal of the Audio Engineering Society}, accepted for
  publication, 2015.

\bibitem{morgan97}
D.~R. Morgan,
\newblock ``A parametric error analysis of the backward integration method for
  reverberation time estimation,''
\newblock {\em The Journal of the Acoustical Society of America}, vol. 101, pp.
  2686--2693, 1997.

\bibitem{chu78}
W.~T. Chu,
\newblock ``Comparison of reverberation measurements using schroeder’s
  impulse method and decay-curve averaging method,''
\newblock {\em The Journal of the Acoustical Society of America}, vol. 63, no.
  5, pp. 1444--1450, 1978.

\bibitem{schuldt13}
C.~Sch{\"u}ldt and P.~H{\"a}ndel,
\newblock ``Blind low-complexity estimation of reverberation time,''
\newblock in {\em Proceedings of the IEEE Workshop on Applications of Signal
  Processing to Audio and Acoustics (WASPAA)}, Oct. 2013.

\bibitem{halmer04}
D.~Halmer, G.~von Basum, P.~Hering, and M.~M{\"u}rtz,
\newblock ``Fast exponential fitting algorithm for real-time instrumental
  use,''
\newblock {\em Review of Scientific Instruments}, vol. 75, no. 6, pp.
  2187--2191, 2004.

\bibitem{jeub09}
M.~Jeub, M.~Schafer, and P.~Vary,
\newblock ``A binaural room impulse response database for the evaluation of
  dereverberation algorithms,''
\newblock in {\em Proceedings of the 16th International Conference on Digital
  Signal Processing}, July 2009.

\end{thebibliography}

\end{document}